\documentclass[aps,prd,reprint,preprintnumbers,superscriptaddress,showpacs,twocolumn]{revtex4-1}
\usepackage{latexsym,graphicx,amssymb,amsmath,mathrsfs}
\usepackage{setspace,bm}
\usepackage[breaklinks, colorlinks=true, pdfstartview=FitV, linkcolor=red, citecolor=blue, urlcolor=blue]{hyperref}

\usepackage{amsmath}
\usepackage[usenames]{color}
\usepackage{latexsym}
\usepackage{epstopdf}

\newcommand{\Slash}[1]{{\ooalign{\hfil/\hfil\crcr$#1$}}}
\newcommand\nc{N_\mathrm{c}}
\newcommand\nf{N_\mathrm{f}}

\newcommand\mur{\mu_\mathrm{R}}
\newcommand\mui{\mu_\mathrm{I}}
\newcommand\trw{T_\mathrm{RW}}

\newcommand\JSD{D_\mathrm{JS}}
\usepackage[normalem]{ulem}  

\newcommand{\comment}[1]{}

\renewcommand\sout{\bgroup \color{red} \ULdepth=-.5ex \ULset}

\begin{document}
\preprint{YITP-17-132}

\title{
Investigation of confinement-deconfinement transition via probability distributions
}

\author{Kouji Kashiwa}
\email[]{kouji.kashiwa@yukawa.kyoto-u.ac.jp}
\affiliation{Yukawa Institute for Theoretical Physics,
Kyoto University, Kyoto 606-8502, Japan}

\author{Akira Ohnishi}
\email[]{ohnishi@yukawa.kyoto-u.ac.jp}
\affiliation{Yukawa Institute for Theoretical Physics,
Kyoto University, Kyoto 606-8502, Japan}

\begin{abstract}
We investigate the confinement-deconfinement transition
 at finite temperature in terms of the probability distribution
 of Polyakov-loop complex-phase
 via the Jensen-Shannon divergence.
 The Jensen-Shannon divergence quantifies
 the difference of two probability
 distributions, namely the target and reference probability
 distributions.
We adopt the complex-phase distributions of the spatially averaged
 Polyakov loop  at $\mu/T=0$ and $\mu/T=i\pi/3$
 as the target and reference distributions, respectively.
It is shown that the Jensen-Shannon divergence has the inflection
 point when the target system approaches the Roberge-Weiss
 endpoint temperature even in the finite-volume system.
This means that we can detect
 the confinement-deconfinement
 transition from the structural change of probability distributions when
 we suitably set the reference probability distribution.
It is also shown that we can pick up the information of the
 confinement-deconfinement transition from the quark
 number density by using the Fourier decomposition;
 Fourier coefficients have a long tail at around the transition temperature
 and show a divergent series in calculating the normalized kurtosis.
\end{abstract}

\maketitle

\section{Introduction}

Understanding nonperturbative properties of Quantum Chromodynamics (QCD)
at finite temperature ($T$) and real quark chemical potential ($\mur$)
is one of the important and interesting subjects in the nuclear and
elementary particle physics.
There are two important and famous non-perturbative phenomena in QCD;
the chiral and confinement-deconfinement transitions.
The chiral phase transition is well understood from the spontaneous
chiral symmetry breaking, but the confinement-deconfinement transition
is rather unclear.
In the pure Yang-Mills limit, the Polyakov loop representing the gauge
invariant holonomy becomes the  exact order-parameter of the
confinement-deconfinement transition.
In comparison, the Polyakov loop is no longer an order parameter of the
confinement-deconfinement transition in the system with dynamical
quarks.
This may mean that we cannot discuss the confinement-deconfinement
transition by using the spontaneous symmetry breaking.

Recently, a new determination of the confinement-deconfinement
transition has been proposed by using the imaginary chemical potential
($\mui$) in the system with dynamical quarks~\cite{Kashiwa:2015tna}.
The main idea is based on an analogy of the topological order discussed
in the condensed matter physics~\cite{Wen:1989iv} and QCD at
$T=0$~\cite{Sato:2007xc} where the ground-state degeneracy is used to
clarify the transition associated with the topological order.
Actually, the non-trivial free-energy degeneracy at
finite $\theta \equiv \mui/T$ is used to replicate the ground-state
degeneracy in Ref.~\cite{Kashiwa:2015tna};
the topological order is based on the topological modification of
the system and thus we introduce the imaginary chemical potential to
modify the system's topological structure.
The lattice QCD action has the Roberge-Weiss (RW) $Z_3$ symmetry,
$U_\nu(x)\to e^{-i2k\pi/N_c}U_\nu(x)$
and
$\theta\to\theta+2k\pi/N_c$ with $U$ and $k$ being the link variable and
an integer, respectively,
and the states before and after the RW transformation can be
interpreted as different states
when non-trivial degeneracy exists.
By using the determination, we can clarify the confinement-deconfinement
transition from topological points of view even if thermodynamics
indicate the crossover behavior; see appendix~\ref{Sec:app} for details.
Then, the quantum order-parameter which is so called the quark number
holonomy can be constructed by considering the
contour-integral of the quark number density along $\theta=0 \sim
2\pi$~\cite{Kashiwa:2016vrl}.
Unfortunately, the quark number holonomy in the
finite size system becomes exactly zero at any $T$
without the extrapolation to infinite volume limit.
Thus, the quark number holonomy is not
easy to obtain in the lattice QCD simulation.

In this article, we propose a new order parameter of the
confinement-deconfinement transition which has better properties than
the quark number holonomy;
we employ the {\it Jensen-Shannon divergence}~\cite{lin1991divergence}
of the Polyakov-loop complex-phase distribution.
The Jensen-Shannon divergence which
is based on the Kullback-Leibler
divergence~\cite{kullback1951information} is defined by using two
probability distributions, namely the reference and target probability
distributions, and it quantifies the difference between them.
Thus, it has been widely used in the several research field such as the
machine learning, statistical dynamics and information theory.
For example in physics, the Kullback-Leibler
divergence provides the important understanding of the mean-field
approximation.
In this article, we try to apply it to detect the confined and deconfined
phases by devising the reference probability distribution.
We demonstrate that the Jensen-Shannon divergence rapidly grows
as a function of $T$ at around the transition temperature.

We also discuss
the quark number density to detect the confinement-deconfinement transition.
Unfortunately, bulk quantities are expected to be insensitive to
topologically determined confinement-deconfinement transition and thus
we employ the Fourier decomposition to pick up non-trivial correlation
between the quark and hadronic dynamics.
By performing the Fourier decomposition of the quark number density
at finite $\theta$,
we find that the Fourier coefficients have a long tail at large baryon numbers.
This long tail leads to a divergent series in the calculation
of the kurtosis, the fourth order cumulant of the net baryon number.

This paper is organized as follows.
Section.~\ref{Sec:CE} shows the Jensen-Shannon divergence and our
setting to clarify the topological confinement-deconfinement transition.
Several discussions are shown in Sec.~\ref{Sec:Discussion} and
Sec.~\ref{Sec:Summary} is devoted to summary.

\section{Jensen-Shannon divergence}
\label{Sec:CE}

In this section, we
briefly explain the Jensen-Shannon divergence which is proposed in
the information theory. We start from the cross entropy which is closely
related with the Shannon entropy.
The cross entropy is defined as
\begin{align}
 H(p,q) &= - \sum_{x} p(x) \ln q(x).
\label{Eq:CE_D}
\end{align}
where $p$ and $q$ are the discrete reference and target probability
distributions, respectively.
In the case with the continuous probability distributions,
Eq.~(\ref{Eq:CE_D}) is turned into the form
\begin{align}
H(p,q) &= - \int p(x) \ln q(x)~dx,
\label{Eq:CE_C}
\end{align}
The cross entropy can be expressed as
\begin{align}
 H(p,q) &= H(p) + D_\mathrm{KL}(p||q),
\end{align}
with
\begin{align}
H(p) &\equiv - \int p(x) \ln p(x)~dx,
\nonumber\\
D_\mathrm{KL}(p||q)
&\equiv \int p(x) \ln \Bigl[ \frac{p(x)}{q(x)}\Bigr]~dx,
\end{align}
where $D_\mathrm{KL}(p||q)$ means the Kullback-Leibler
divergence and $H(p)$ is the entropy for $p$.

Since the probability distributions must be positive, the
Kullback-Leibler divergence should be positive for any $p$ and
$q$.
In the case with $p(x)=q(x)$, we have $D_\mathrm{KL}(p||p)=0$ which is
the minimal value of the Kullback-Leibler divergence, namely the minimum
principle.
It should be noted that the exchange of $p$ and $q$ leads
to an asymmetric relation, $D_\mathrm{KL}(p||q) \neq D_\mathrm{KL}(q||p)$.
In addition, the Kullback-Leibler divergence is not bounded above.
Because of these undesirable properties, we use the Jensen-Shannon
entropy below.

The Jensen-Shannon divergence is an extended quantity of the Kullback-Leibler
divergence which is symmetric against exchange of $p$ and $q$ and also
bounded above.
The definition of the Jensen-Shannon divergence
($\JSD$) is
\begin{align}
\JSD
&= \frac{1}{2} \Bigl[ D_\mathrm{KL} (p||M) + D_\mathrm{KL}(q||M) \Bigr],
\end{align}
where $M = (p+q)/2$.
This quantity is symmetric because of the structure of $M$ and exists in
the range;
\begin{align}
0 \le \JSD \le \ln 2.
\end{align}
$\JSD$ is zero when two distributions are the same,
and becomes $\ln 2$ when two distributions have no overlap.
Because of these good properties, we employ the Jensen-Shannon entropy
instead of the cross entropy and the Kullback-Leibler divergence in
this article.

\subsection{Application of Jensen-Shannon divergence}

In this study, we set the target and reference
probability distributions at $\mu/T=0$ and
$\mu/T =i\pi/3$ at the same $T$, respectively.
The actual probability distribution may be prepared by using the
histogram of the spatially averaged Polyakov-loop phase ($\phi$)
in the lattice QCD simulation;
\begin{align}
 p(\phi) &= \frac{n(\phi;\theta=0)}{n_\mathrm{conf}},
\\
 q (\phi)
 &= \lim_{\epsilon \to 0}
 \frac{n(\phi-\theta;\theta = \frac{\pi}{3})}{n_\mathrm{conf}},
\label{Eq:ref_prob}
\end{align}
where $n(\phi;\theta)$ means the number of the
configurations at $\phi$ with $\theta$ and $n_\mathrm{conf}$ are total
number of configurations.
In Eq.~(\ref{Eq:ref_prob}), we take into account the phase shift of
$\phi$ by $-\theta$ because there is the trivial phase shift of
the probability distribution and it is
not desirable in this study because we just wish to compare the shape
difference of the target and reference probability distributions.
Intuitively, the structural change of $q$ are induced by
the appearance of sufficiently strong quark-gluon dynamics comparing with hadronic
dynamics in the thermodynamic system.
Thus, it is natural
to think that the structural change of $q$ reflects the confinement and
deconfinement dynamics of QCD.
Actually, this is the indirect support for Ref.~\cite{Kashiwa:2015tna}.

In the actual lattice QCD calculation, we cannot have infinite number
of the configurations and thus we may fit the histogram for finite
number of configurations to obtain the continuous probability
distributions by the analytic function.
It should be noted that we must consider bins of $\phi$
to construct the histogram for the discrete distributions before the
fitting and then the fitted histogram depends on the bin size.
This procedure is nothing but the coarse graining.
Therefore, we should use the same bin size for all calculations.
It should be noted that there is another choice to generate the
distributions;
we can calculate the Polyakov loop on each lattice site
and thus we can construct the histogram as a function of $\phi$
without the spatially averaging.
In this case, statistics may be improved because the number of
configurations is effectively enhanced.

The extension of the present Jensen-Shannon divergence to the finite
density is straightforward; we should introduce the same value of the real
chemical potential ($\mur$) to the
reference and target probability distributions.
Therefore, we do not show explicit form of the Jensen-Shannon
divergence at finite $\mur$ in this article.
From the viewpoint of the extension, the Jensen-Shannon divergence
is better quantity than the quark number
holonomy because we do not need the complex chemical potential in the
calculation of the Jensen-Shannon divergence at finite $\mur$.

\subsection{Gaussian distribution}

To discuss the qualitative behavior of the Jensen-Shannon divergence,
we start from the Gaussian distribution which is the
ideal situation where probability distributions have one peak
and are localized.
We can expect different type of the probability
distribution in QCD
such as the two or more peak structural distribution near the phase
transition which are discussed in the next subsection.

The probability distributions are limited in the $ -\pi \le \phi < \pi$
region in our case, but we may approximate these by the
Gaussian distribution when the variance is sufficiently small.
The continuous Gaussian distribution is
\begin{align}
 f(x;\sigma^2,x_0)
 &= \frac{1}{\sqrt{2 \pi \sigma^2}}
 \exp \Bigl[ - \frac{(x-x_0)^2}{2 \sigma^2} \Bigr],
\end{align}
where $\sigma^2$ is the variance and $x_0$ is the expectation value of $x$.
If we set $p = f(\phi;\sigma_p^2,\phi_p)$ and $q = f(\phi; \sigma_q^2,\phi_q)$,
the Kullback-Leibler divergence becomes
\begin{align}
 D_\mathrm{KL} (p||q)
 &= \frac{1}{2} \Bigl[
    \ln \Bigl( \frac{\sigma_q^2}{\sigma_p^2} \Bigr)
    + \frac{\sigma_p^2+(\phi_p-\phi_q)^2}{\sigma_q^2} - 1
    \Bigr].
 \label{Eq:CE-si}
\end{align}
If we consider corrections for the approximation, Eq.~(\ref{Eq:CE-si})
should contain the exponential and error functions.
We can clearly see that
the variance strongly affects the Kullback-Leibler divergence
and equivalently the Jensen-Shannon divergence.
If the configurations are well localized and its probability
distribution has
one peak, the Kullback-Leibler divergence obeys Eq.~(\ref{Eq:CE-si}).
In the lattice QCD simulation, it may be happen if the artifact of the
hybrid Monte-Carlo method provides the localized one-peak probability
distributions in the case with the first-order transition.

When we can correctly generate configurations
in the case with the first-order transition, the
Kullback-Leibler divergence behaves differently from the simple Gaussian
distribution case.
The phase transition is smeared by the finite size effect, but the
remnant of the phase transition survives in the structure of the
effective potential and thus the variance and mean value can feel the
smeared phase transition.
For example, the spreading tendency of configurations in the $\phi$-axis
at the RW transition can be found in Fig.~1 of Ref.~\cite{Bonati:2010gi}.
We demonstrate it in the next subsection.

\subsection{Model distribution in QCD}

In this section, we estimate the asymptotic behavior of the
Jensen-Shannon divergence and calculate it by using a QCD effective
model, numerically.

\subsubsection{Estimation}
\label{Sec:estimation}

At low $T$, we can expect that the target and reference probability
distributions ($p$ and $q$) have
the one-peak structure with the same peak positions.
Then, the Jensen-Shannon divergence can become small when the difference
between the variances are small;
the Jensen-Shannon entropy is only written in terms of the ratio,
$\sigma_p^2 / \sigma_q^2$.
Therefore, we can never have $\ln 2$ at low $T$.

At the temperature above the RW endpoint,
$q$ has two peaks,
since the two states related by the RW transformation,
$\phi\simeq 0$ and $\phi\simeq -2\pi/3$,,
have the same free-energy at $\theta=\pi/3$.
If we can approximate the two peak structure by summing up two
Gaussian distributions ($f_1$ and $f_2$),
the peak position of $f_1$ and $f_2$ is departed from $\phi=0$ and
then the mean values joins the calculation.
When the system volume increases, the variance of $f_1$ and $f_2$ becomes
small and thus
the second term of Eq.~(\ref{Eq:CE-si}) becomes the dominant
contribution.
Then, the overlap between $p$ and $q$ should be small.
In the zero-overlap limit of $p$ and $q$, the Jensen-Shannon divergence
becomes $\ln 2$ and thus it should be the asymptotic value.
Therefore, we can expect the $T$-dependence of the Jensen-Shannon
divergence as $0 \to \ln 2$ with increasing $T$.
In the infinite volume limit, the Jensen-Shannon entropy becomes $\ln 2$
above RW endpoint and this behavior is universal for the second and first
order RW endpoints.

\subsubsection{Demonstration}
To demonstrate the situation with the smeared first-order transition in
the finite volume system, we here use the Polyakov-loop quark (PQ) model;
this model is corresponding to the Polyakov-loop extended
Nambu--Jona-Lasinio (PNJL) model~\cite{Fukushima:2003fw} without the
chiral symmetry breaking.
The PQ model can reproduce the RW periodicity and its
transition and thus it is enough to use in this study.

The Lagrangian density of the PQ model is
\begin{align}
{\cal L} &= {\bar q} \Slash{D} q + {\cal V}_{\mathrm{gluon}} (\Phi,{\bar \Phi}),
\end{align}
where the covariant derivative is
$D_\nu = \partial_\nu  - i g A_\nu \delta_{\nu 4}$,
$\Phi$ (${\bar \Phi}$) denotes the Polyakov loop (its conjugate) and
${\cal V}_{\mathrm{gluon}}$ expresses the gluonic contribution.
With the mean-field approximation, the effective potential is expressed
as
\begin{align}
{\cal V} &= {\cal V}_\mathrm{quark} + {\cal V}_\mathrm{gluon},
\end{align}
where ${\cal V}_\mathrm{quark}$ is the quark contribution.
The actual form of ${\cal V}_\mathrm{quark}$ becomes
\begin{align}
{\cal V}_\mathrm{quark}
&= - 2 \nf \int \frac{d^3 p}{(2\pi)^3}
   \Bigl[ T \ln \Bigl( f^- f^+ \Bigr)\Bigr],
\end{align}
where we ignore the $\Phi$-independent term and
\begin{align}
f^-
&= 1
 + 3 (\Phi+{\bar \Phi} e^{-\beta E_{\bf p}^-} ) e^{-\beta E_{\bf p}^-}
 + e^{-3\beta E_{\bf p}^-},
\nonumber\\
f^+
&= 1
 + 3 ({\bar \Phi}+\Phi e^{-\beta E_{\bf p}^+} ) e^{-\beta E_{\bf p}^+}
 + e^{-3\beta E_{\bf p}^+},
\end{align}
with $E_{\bf p}^\mp = E_{\bf p} \mp \mu = \sqrt{{\bf p}^2 + M^2} \mp \mu$.
In this paper, we choose the polynomial Polyakov-loop
potential~\cite{Ratti:2005jh} as the gluonic contribution;
\begin{align}
{\cal V}_\mathrm{gluon}
 &= T^4 \Bigl[- \frac{b_T}{2} \Phi {\bar \Phi}
              - \frac{b_3}{6} (\Phi^3 + {\bar \Phi}^3)
              + \frac{b_4}{4} (\Phi {\bar \Phi})^2 \Bigr],
\end{align}
where
\begin{align}
 b_T
 &= a_0 + a_1 \Bigl(\frac{T_0}{T} \Bigr)
        + a_2 \Bigl(\frac{T_0}{T} \Bigr)^2
        + a_3 \Bigl(\frac{T_0}{T} \Bigr)^3,
\end{align}
because this form does not have the singularity unlike the
logarithmic Polyakov-loop potential and thus it is
convenient to prepare the probability distributions.
The parameters are set to reproduce the lattice QCD data
in the pure Yang-Mills limit as
\begin{align}
 a_0&=6.75,~~a_1=-1.95,~~a_2=2.62,~~a_3=-7.44,
 \nonumber\\
 b_3&=0.75,~~b_4=7.5,
\end{align}
with $T_0 = 270$ MeV.
This Polyakov-loop potential leads to the second-order RW
endpoint.

We construct the target and reference probability distributions by
integrating out the absolute value of the Polyakov loop ($R$) as
\begin{align}
prob.~dist. =
 \frac{ \int e^{-\frac{V}{T}{\cal V}} d R} {\int e^{-\frac{V}{T}{\cal V}} d R~d \phi},
\end{align}
where $V$ means the
three-dimensional volume, $\phi$ is the phase of the Polyakov loop.
In this paper, we estimate ${\cal V}$ in the thermodynamic limit and
take finite $V$ in the calculation of the Jensen-Shannon divergence to
mimic the finite size system.

\begin{figure*}[t]
 \centering
 \includegraphics[width=0.4\textwidth]{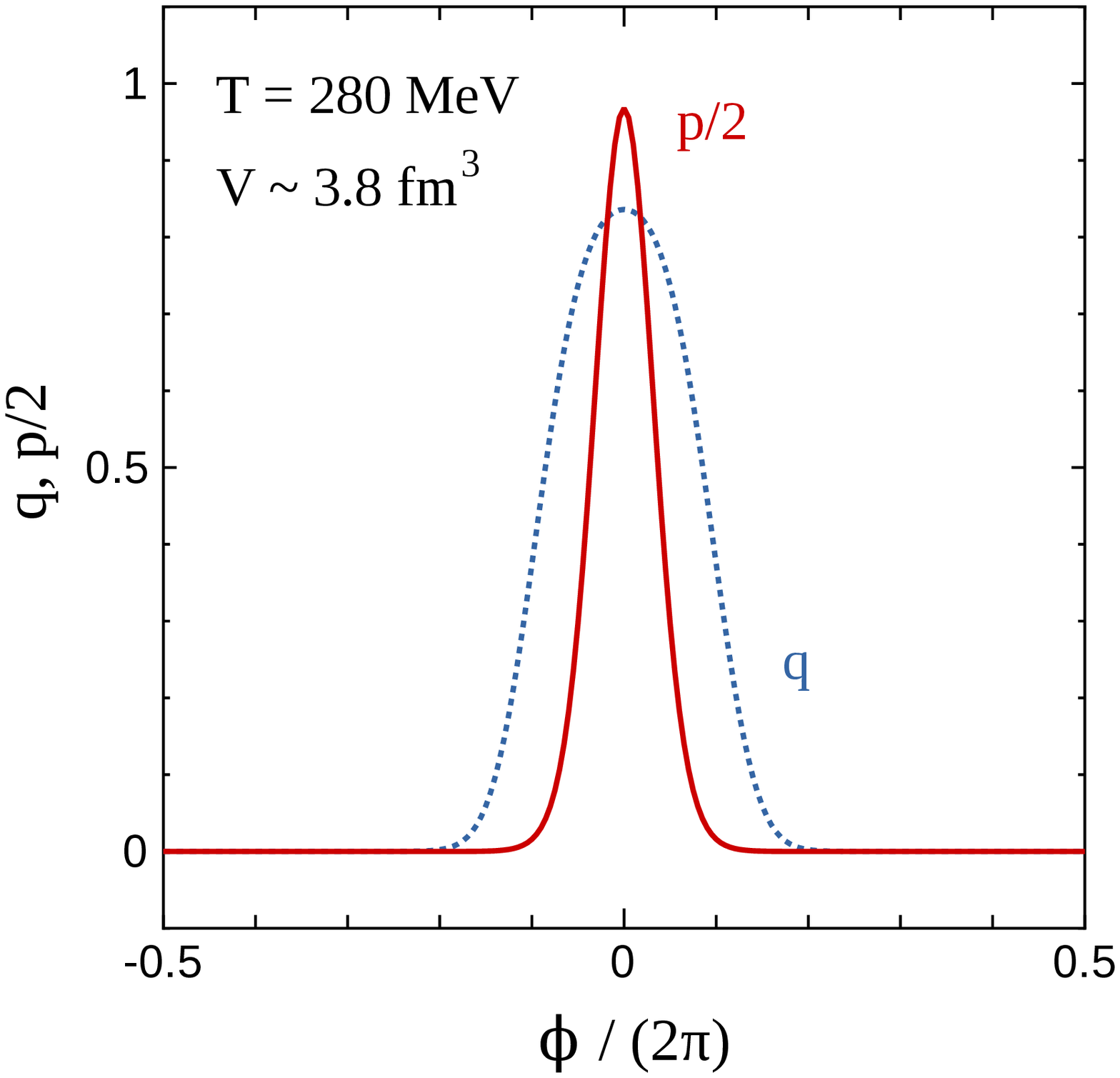}
 \hspace{18mm}
 \includegraphics[width=0.4\textwidth]{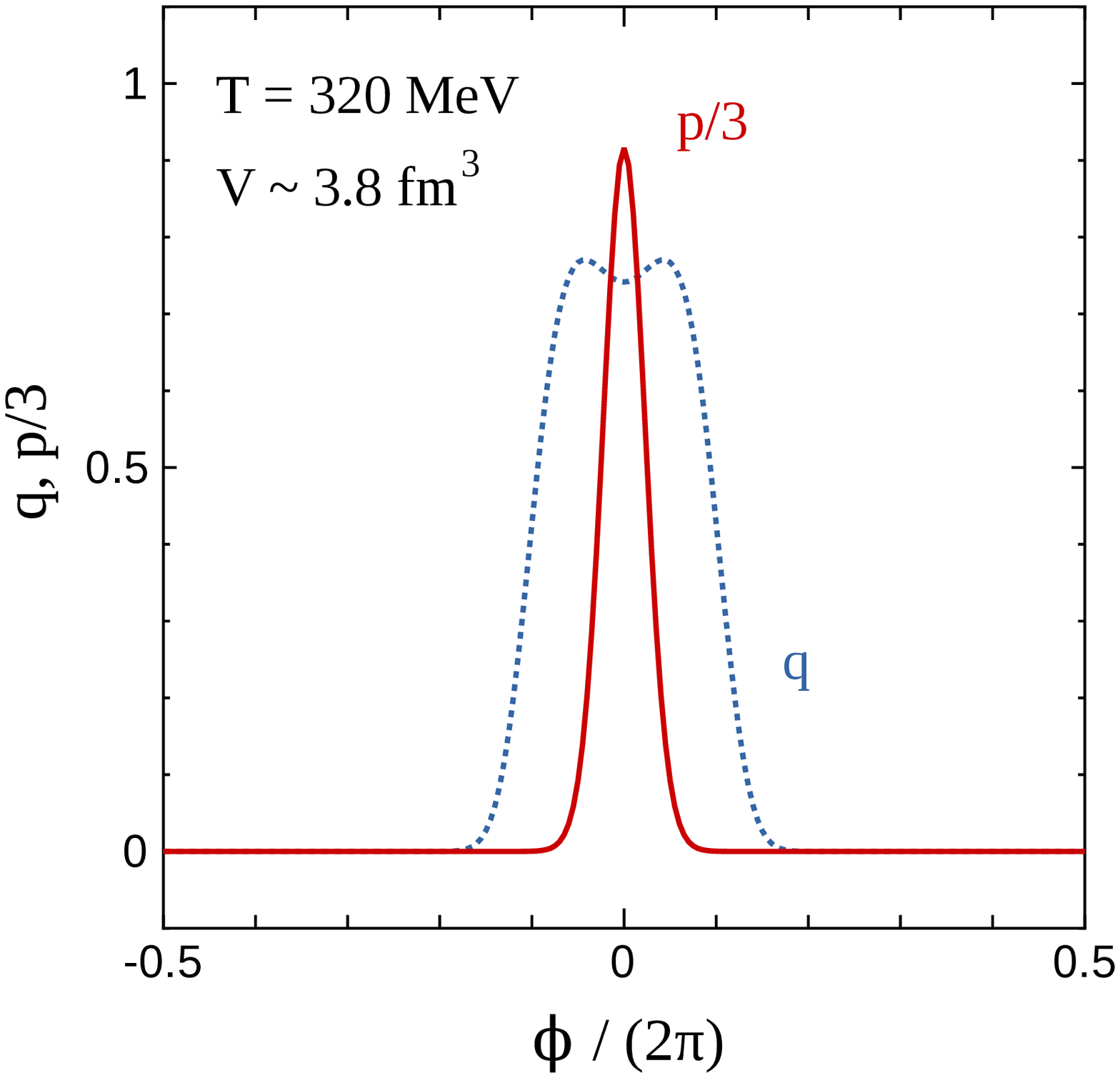}
 \caption{
 The left (right) panel shows the probability distributions at $T=280$
 ($320$) MeV with $V\sim 3.8$ fm$^3$ in the PQ model as a function of
 $\phi/(2\pi)$.
 }
\label{Fig:PD}
\end{figure*}

Firstly, we show the probability distributions in the PQ model at $V =
500$ GeV$^{-3}$ $\sim 3.8$
fm$^3$ in Fig.~\ref{Fig:PD}.
At low $T$, the overlap between the probability distributions are large;
the Jensen-Shannon divergence should be small.
In comparison, the overlap between $p$ and $q$ are small at high $T$ and
thus Jensen-Shannon divergence should be large.

\begin{figure}[h]
 \centering
 \includegraphics[width=0.45\textwidth]{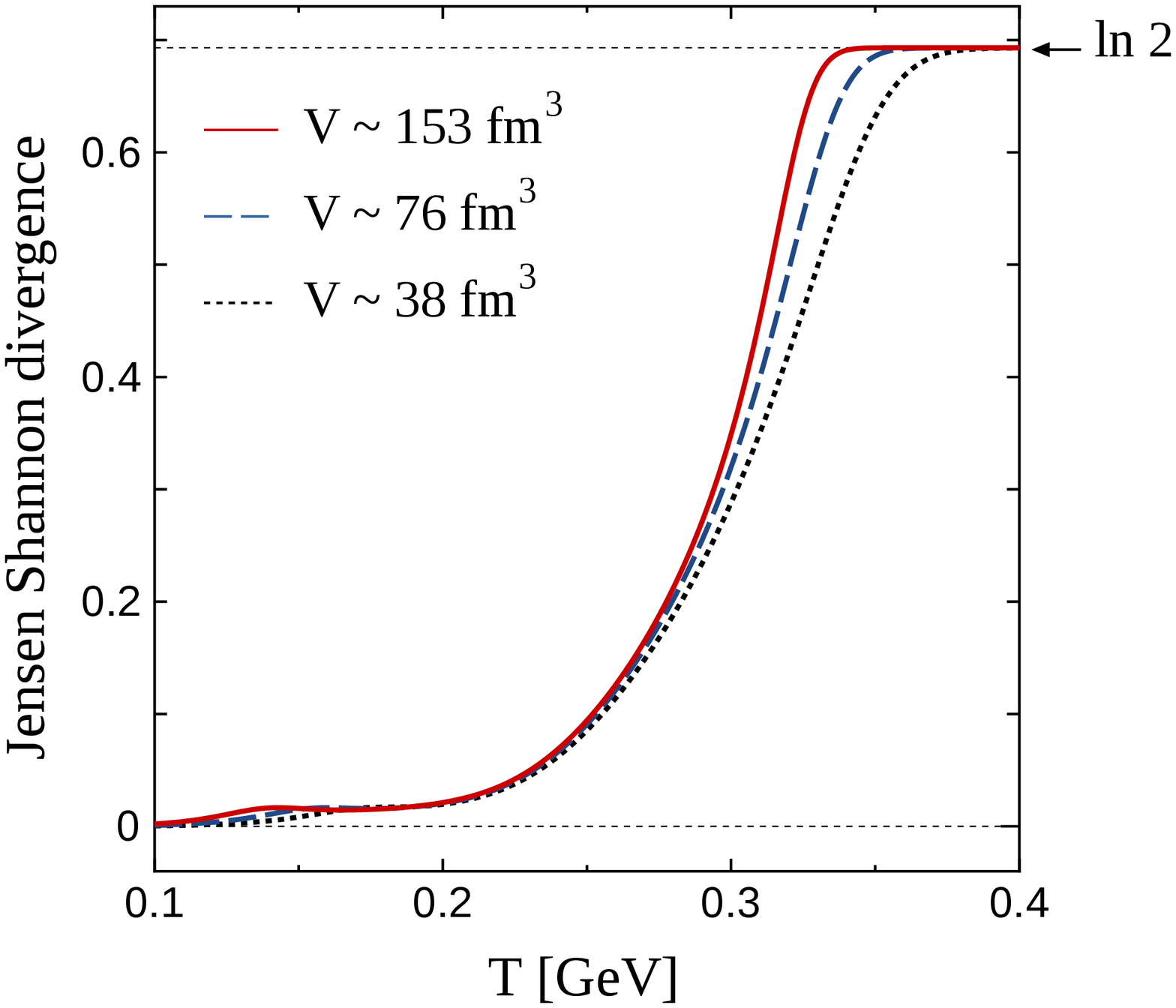}
 \\
 \vspace{0.5cm}
 \hspace{-1cm}
 \includegraphics[width=0.41\textwidth]{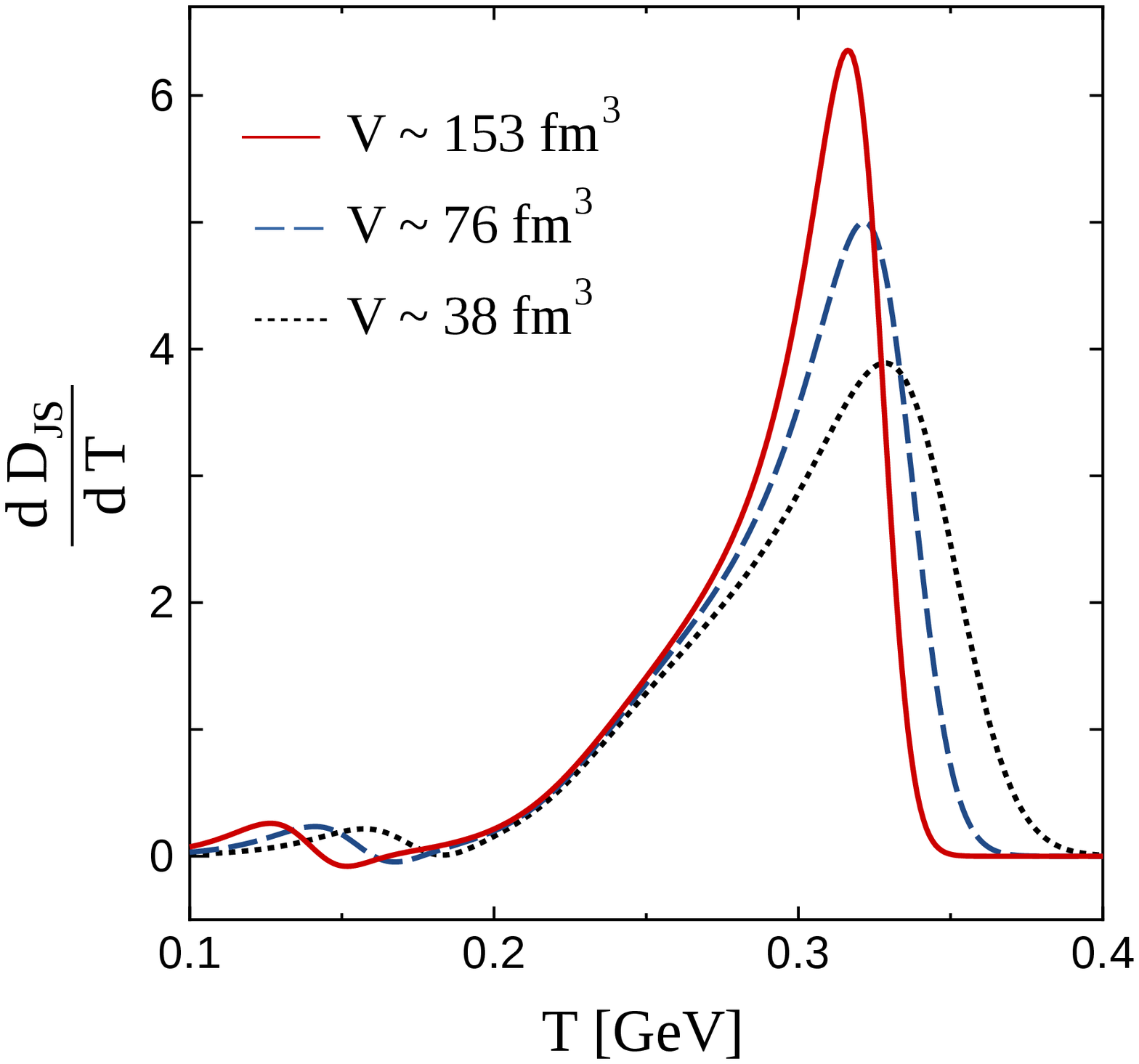}
 \caption{
 The top panel shows the Jensen-Shannon divergence as a function of $T$ in
 the PQ model and the bottom one does its derivative by $T$.
 The dotted, dashed and solid lines represent the
 results with $V\sim 38$, $76$ and $153$ fm$^3$,
 respectively.
 }
\label{Fig:JSD-PQ}
\end{figure}

The top panel of Fig.~\ref{Fig:JSD-PQ} shows the Jensen-Shannon
divergence as a function of $T$ in the PQ model.
We can clearly see that it approaches to $\ln 2$ as estimated in
Sec.~\ref{Sec:estimation} around the RW endpoint temperature;
$T_\mathrm{RW}$ is about $300$ MeV in the present model.
Also, we can see that
the volume dependence strongly appears in the deconfined phase.
The bottom panel of Fig.~\ref{Fig:JSD-PQ} shows the $T$-derivative of
the Jensen-Shannon divergence.
We can see that the peak position of $d \JSD/dT$ approaches to
$T_\mathrm{RW}$ with increasing $V$.
When we determine the pseudo-critical temperature from $d \Phi/ dT$, the
pseudo-critical temperature of the
confinement-deconfinement transition is about $230$ MeV
and it is much lower than $T_\mathrm{RW}$.
This value is irrelevant to
the topological phase transition.
Actually, the Polyakov loop only feels the target probability
distribution and thus this result seems to be trivial.

If we interpret the present Jensen-Shannon divergence as the
information blocking process along $\theta$ by the thermal system,
we can say that the deconfined phase strongly blocks the information
propagation than that in the confined phase.
In other words, the target probability distribution is not very sensitive
to changing of degree of freedoms by the thermal bath.

\section{Discussion}
\label{Sec:Discussion}

In this section, we investigate the possibility to detect the
topologically determined confinement-deconfinement transition from the
quark number density because it is more physicist-friendly quantity than
the Jensen-Shannon divergence.

\begin{figure}[t]
 \centering
 \includegraphics[width=0.5\textwidth]{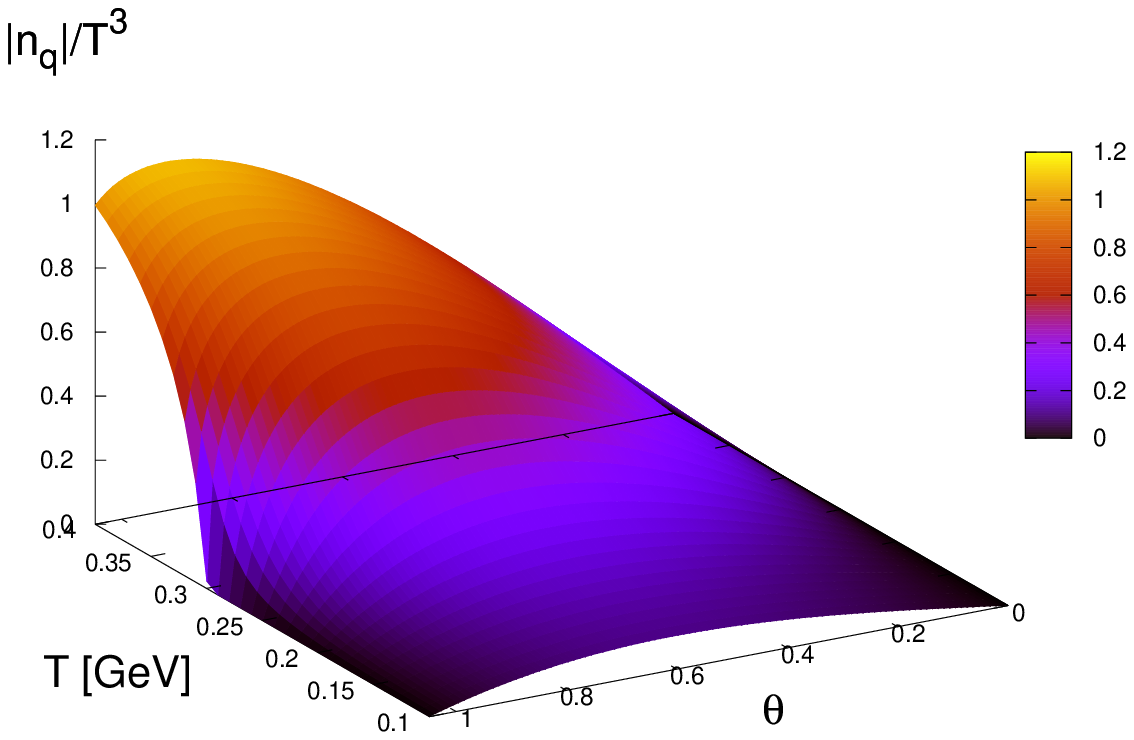}
 \caption{The absolute value of the quark number density in the
 $T$-$\theta$ plane.
 }
\label{Fig:Tnq}
\end{figure}

Bulk quantities such as  the pressure and the quark number density are
expected to be insensitive to the topologically determined
confinement-deconfinement transition, but we may see the information
from these quantities by employing the Fourier decomposition because it
can pick up the structural change at finite $\theta$.
Actually, we here employ the quark number density and
the actual behavior of it in the PQ model in the $T$-$\theta$ plane is shown in
Fig.~\ref{Fig:Tnq}.

\begin{figure}[t]
 \centering
 \includegraphics[width=0.4\textwidth]{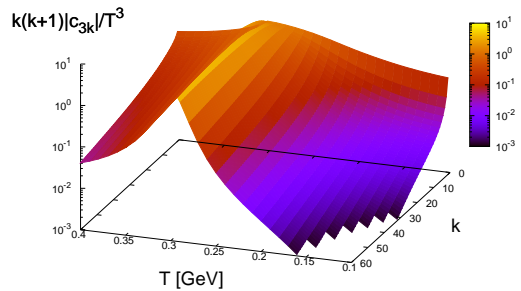}
 \caption{
 Fourier coefficients in the $T$-$k$ plane.
 The coefficients are normalized by $k(k+1)/T^3$.
 }
\label{Fig:c3k}
\end{figure}

The quark number density is fitted by using $\theta$-odd periodic
functions such as
\begin{align}
 F_\mathrm{fit}
 &= c_1(T) \sin\theta + \sum_{k} c_{\nc k}(T) \sin ( \nc k \theta),
\label{Eq:c3k}
\end{align}
where $c$ are fitting parameters.
Since
there is the RW transition at $\theta=\pi/3$,
we restrict the fitting region
as $0 \le \theta \le \pi/3-0$.
This fit is nothing but the Fourier decomposition.
Since the quark number density is a $2\pi/N_c$-periodic function
of $\theta$,
$\sin(N_ck\theta)$ terms should be enough in the Fourier series in principle,
but the discontinuity at $\theta=\pi/N_c$ makes the convergence slower.
Here, we introduce the one-quark-like contribution which is the first
term in Eq.~(\ref{Eq:c3k}) and sufficient
number of hadronic contributions which is the second term of
Eq.~(\ref{Eq:c3k}).
By introducing the first term with
$c_1=\mathrm{Im}(n_q)(\theta)/\sin\theta|_{\theta \to \pi/N_c-0}$,
we can remove the discontinuity at $\theta=\pi/N_c$.
The Fourier coefficients in the $T$-$k$ plane is shown in
Fig.~\ref{Fig:c3k}.
The Fourier coefficients, $c_{\nc k}=c_{3 k}$, rapidly decrease at $T <
\trw$, but its behavior is slow at $T>\trw$.
Also, there is the peak structure of $c_{\nc k}$
around $T=\trw$.
The actual behavior of $c_{3 k}$ around $\trw$
can be approximated as
\begin{align}
c_{3k} \sim \frac{(-1)^{k-1}}{k(k+1)},
\end{align}
with some exponential factors depending on $T$.
This behavior of $c_{3k}$ can be judged to be long-tailed.
For example, the Fourier coefficients in the free massless baryons
behave as $c_{3k}\propto (-1)^{k-1}/k^3$.
The long tail suggests that the contribution of other Fourier components,
such as $\sin2\theta$.

There are some implications of the Fourier coefficients in
observables such as the normalized
kurtosis~\cite{Ejiri:2005wq,Karsch:2010ck}.
The definition at $\mu=0$ is
\begin{align}
\kappa \sigma^2 &= \chi_\mu^{(4)}/\chi_\mu^{(2)},
\end{align}
where
\begin{align}
\chi_\mu^{(2)}
&= \frac{1}{VT^3} \frac{\partial^2 \ln {\cal Z}}{\partial (\mu/T)^2}
 = \frac{1}{T^3} \frac{\partial n_q}{\partial (\mu/T)}
 = \frac{1}{T^3} \frac{\partial~ \mathrm{Im} (n_q)}{\partial \theta},
\nonumber\\
\chi_\mu^{(4)}
&= \frac{1}{VT^3} \frac{\partial^4 \ln {\cal Z}}{\partial (\mu/T)^4}
 = - \frac{1}{T^3} \frac{\partial^3~ \mathrm{Im}(n_q)}{\partial \theta^3}.
\end{align}
We can easily understand that there is the region where the Fourier
series fails to converge in calculating
the normalized kurtosis
around $T=\trw$ from the behavior of $c_{3k}$,
\begin{align}
\chi_\mu^{(4)}=\frac{c_1}{T^3}+\sum_k \frac{N_c^3 k^3 c_{3k}}{T^3}\ .
\end{align}
If we calculate the normalized kurtosis in the present way, we may need
the resummation of the $c_{3k}$ and the actual resummation will be
discussed elsewhere.
This failure behavior indicates non-trivial correlations between
quark and hadronic dynamics.
Therefore, we can access the topologically determined
confinement-deconfinement
transition from the behavior of Fourier coefficient
even if we can not find any signals from bulk quantities.

\section{Summary}
\label{Sec:Summary}

In this study, we have considered the Jensen-Shannon divergence as
the new order-parameter of the topological confinement-deconfinement
transition at finite temperature ($T$).
The Jensen-Shannon divergence was proposed in the information theory
and it can detect the difference between two probability
distributions, namely the reference and target probability distributions.

For the determination of the topological confinement-deconfinement
transition, we set the
target and reference probability distributions at $\mu/T =0$ and
$\mu/T =i\pi/3$ with the same $T$.
Actually, we consider the histogram of configurations in the lattice QCD
simulation as a function of the Polyakov-loop phase as the probability
distributions.
To compare the shape difference between the probability distributions,
we shift the phase of the reference probability distribution to remove
the trivial rotation of the probability distribution by $\theta$.
In the actual calculation, it is convenient to use the fitting of the
histogram by analytic functions because we cannot gather the
infinite number of configurations.
Then, the coarse graining should be cared because the value of the
Jensen-Shannon divergence depends on it.
With these treatments, we can calculate the continuous form of the
Jensen-Shannon divergence.

Even in the case with the smeared first-order transition,
the variance and the mean value of the probability distributions can feel
the transition existing in the infinite volume system.
From the estimation by using the Gaussian distribution, we can expect
that the Jensen-Shannon divergence becomes small in the confined phase
and it approaches to $\ln 2$ in the deconfined phase.
We demonstrate it by using the Polyakov-loop quark model.
The numerical results exactly support the simple estimation and then we
can detect the topological confinement-deconfinement transition from the
Jensen-Shannon divergence.

To explore the possibility to detect the topologically determined
confinement-deconfinement transition,
we investigate the Fourier decomposition of the quark number density
and discuss the normalized kurtosis.
We found that the Fourier coefficients ($c_{3k}$) has peak around the RW
endpoint temperature.
The behavior of $c_{3k}$ indicates that
there is the region where the normalized kurtosis
calculated from the Fourier coefficients fails to converge.
We can consider that
this behavior shows how strong the quark and hadronic dynamics are
non-trivially correlated in the thermal system.

Advantages of the present Jensen-Shannon divergence are following:
We can calculate it even in the finite size system
because the variance and the mean value of the
probability distribution can feel
the smeared phase transition in the finite volume system.
Also, we can easily introduce the fluctuation around
the global minima of the effective potential even in the
effective model computations.
This philosophy is similar to the study with the Polyakov-loop
fluctuation~\cite{Lo:2013hla}.
In usual discussions of the confinement-deconfinement transition, we
concentrate on the mean value of expectation values, but the present
study sheds light on the structure of the probability distributions
itself via the Jensen-Shannon divergence and shows the importance of
the selection of the suitable reference probability distribution.

\begin{acknowledgments}
 This work
 is supported in part by the Grants-in-Aid for Scientific Research
 from JSPS (Nos. 15K05079, 15H03663, 16K05350),
the Grants-in-Aid for Scientific
 Research on Innovative Areas from MEXT (Nos. 24105001, 24105008),
 and by the Yukawa International Program for Quark-hadron
 Sciences (YIPQS).
\end{acknowledgments}


\section{Topologically determined confinement-deconfinement transition}
\label{Sec:app}

The famous topological properties of QCD at finite $\mui$ are the
Roberge-Weiss (RW) periodicity and its transition~\cite{Roberge:1986mm}.
The RW periodicity is the special $2\pi/\nc$ periodicity of the grand-canonical
partition function (${\cal Z}$) as a function of $\theta \equiv \mui/T$;
\begin{align}
 {\cal Z} (\theta) &= {\cal Z} \Bigl( \theta + \frac{2 \pi k}{\nc} \Bigr),
\end{align}
where $k \in \mathbb{Z}$.
Existence of the RW periodicity and its properties have been
energetically explored in
the lattice QCD
simulations~\cite{D'Elia:2002gd,deForcrand:2002ci,deForcrand:2003hx,D'Elia:2004at,Chen:2004tb,Bonati:2010gi,Nagata:2011yf,Bonati:2014kpa,Takahashi:2014rta,Doi:2017dyc}.
The RW periodicity appears in all $T$ region, but the origins are
different at low and high $T$ regions.
This difference relates to the first-order RW transition line exists at high
$T$ region; it appears along the $T$-axis at $\theta = (2k-1)\pi/\nc$.
This first-order RW transition line is ended at certain temperature
which is so called the RW endpoint temperature ($\trw$).
At low $T$, the grand-canonical partition function and some other thermodynamic
quantities are smoothly oscillating along the $\theta$-direction, but
these have non-analyticity at high $T$.
This non-analyticity characterizes the RW transition.
Actually, the quark number density has the gap at $\theta=(2k-1)\pi/\nc$
in the high $T$ region and then we can use it to clarify the confined and
deconfined phase based on the analogy of the topological order, namely
the non-trivial free-energy degeneracy~\cite{Kashiwa:2015tna}.

The points $\theta = (2k-1)\pi/3$ where the RW transition appears are
considered as the most sensitive region for changing of the degree of
freedoms in the system, namely which degree of freedoms, quarks and
hadrons, has the supremacy.
It should be noted that $\theta$ does not introduce the additional
energy scale to the system unlike the external magnetic field and $\mur$
and thus it is suitable quantity to investigate the response of the
system.
Similar procedure is possible in the condensed matter physics such as
the one-dimensional quantum wired system bridged on two superconductors
with different phases of the superconducting
gap~\cite{1063-7869-44-10S-S29};
the origin of the periodicity of the energy band along the
difference of the superconducting-gap phase ($\theta_{sc}$)
are different in the topologically trivial and nontrivial phases.
In this case, $\theta_{sc}$ does not introduce the additional energy scale
and also there is the RW-like periodicity.
This indicates that we can investigate the topological phase transition
by investigating the system response against the periodic external
variables if it does not introduce the additional energy scale to the
system.
Our conjecture of the determination of the confinement-deconfinement
transition is based on it.

In the system with dynamical quarks, the Polyakov loop is no longer
the order parameter of the confinement-deconfinement transition and thus
we propose the {\it quark number holonomy} to determine the
transition based on the different realization of the RW
periodicity~\cite{Kashiwa:2016vrl}.
The functional form of the quark number holonomy is defined as
\begin{align}
\Psi
&= \int_{-\pi}^{\pi} \mathrm{Im}
   \Bigl(\frac{\partial {\tilde n}_q}{\partial \theta} \Bigr|_T \Bigr)
   ~d\theta,
\end{align}
where ${\tilde n_q}$ is dimensionless quark number density such as $n_q/T^3$.
The quark number holonomy can count gapped points in the region
$-\pi \le \theta \le \pi$.
Thus, it is characterized by the nontrivial
free-energy degeneracy because the quark number density should have the
gap when the free-energy is non-trivially degenerated.

\bibliography{ref.bib}

\end{document}